\documentclass[12pt]{article}

\begin{document}
\thispagestyle{empty} \vspace*{1cm}
\begin{center}
\vspace{2cm} {\LARGE Schr\"{o}dinger-Newton wave mechanics. \\The
model}

\bigskip
\bigskip
Janina Marciak-Kozlowska\\
Miroslaw Kozlowski\footnote{Corresponding author, e-mail:
MiroslawKozlowski@aster.pl}
\end{center}
\bigskip
Institute of Electron Technology, Al. Lotnik\'ow 32/46, 02-668
Warsaw, Poland

\vspace{2cm}
\begin{abstract}
In this paper the Schr\"{o}dinger equation (SE) with gravity term
is developed and discussed. It is shown that the modified SE is
valid for particles with mass $m<M_P$, $M_P$ is the Planck mass,
and contains the part which, we argue, describes the \textit{pilot
wave}. For $m\to M_P$ the modified SE has the solution with
oscillatory term, i.e. \textit{strings}.
\end{abstract}

\newpage
\section*{Introduction}
When M. Planck made the first quantum discovery he noted an interesting fact~\cite{1}.
The speed of light, Newton's gravity constant and Planck's constant clearly reflect fundamental properties of the world.
From them it is possible to derive the characteristic mass $M_P$, length $L_P$ and time $T_P$ with approximate values
\begin{eqnarray*}
L_P&=&10^{-35} {\rm m}\\
T_P&=&10^{-43} {\rm s}\\
M_P&=&10^{-5} {\rm g}
\end{eqnarray*}
Nowadays much of cosmology is concerned with ``interface'' of gravity and quantum mechanics.

After the \textit{Alpha} moment -- the spark in eternity~\cite{1}
the space and time were created by ``Intelligent Design''~\cite{2}
at $t=T_P$. The enormous efforts of the physicists, mathematicians
and philosophers investigate the \textit{Alpha} moment. Scholars
seriously discuss the \textit{Alpha} moment - by all possible
means: theological and physico-mathematical with growing
complexity of theories. The most important result of these
investigation is the anthropic principle and Intelligent Design
theory (ID).

In this paper we investigate the very simple question: how gravity
can modify the quantum mechanics, i.e. the nonrelativistic
Schr\"{o}dinger equation~(SE). We argue that SE with relaxation
term describes properly the quantum behaviour of particle with
mass $m<M_P$ and contains the part which can be interpreted as the
pilot wave equation. For $m\to M_P$ the solution of the SE
represent the \textit{strings} with mass $M_P$.

\section{Hyperbolic diffusion}
\subsection{Generalized Fourier law}
The thermal history of the system (heated gas container,
semiconductor or Universe) can be described by the generalized
Fourier equation~\cite{3}-\cite{5}
    \begin{equation}
    q(t)=\underbrace{\int_{-\infty}^tK(t-t') }_{\rm thermal \, history}
    \underbrace{\nabla T(t')_{\rule{0pt}{2.7ex}}dt'}_{\rm diffusion}.\label{eq1}
    \end{equation}
In Eq.~(\ref{eq1}) $q(t)$ is the density of the energy flux, $T$
is the temperature of the system and $K(t-t')$ is the thermal
memory of the system
    \begin{equation}
    K(t-t')=\frac{K}{\tau}\exp\left[-\frac{(t-t')}{\tau}\right],\label{eq2}
    \end{equation}
where $K$ is constant, and $\tau$ denotes the relaxation time.

As was shown in~\cite{3}-\cite{5}
    \begin{eqnarray*}
    K(t-t')=
    \left\{\begin{array}{ll}K\lim_{t_0\to 0}\delta(t-t'-t_0)\quad &{\rm diffusion}\\
    K={\rm constant} \quad & {\rm wave}\\
    \frac{K}{\tau}\exp\left[-\frac{(t-t')}{\tau}\right] \quad &{\rm damped \,wave \,or \, hyperbolic\, diffusion}.
    \end{array}\right.
    \end{eqnarray*}
The damped wave or hyperbolic diffusion equation can be written as:
    \begin{equation}
    \frac{\partial^2 T}{\partial t^2}+\frac{1}{\tau}\frac{\partial T}{\partial  t}=\frac{D_T}{\tau}\nabla^2T.\label{eq3}
    \end{equation}
For $\tau\to 0$, Eq.~(\ref{eq3}) is the Fourier thermal equation
    \begin{equation}
    \frac{\partial T}{\partial t}=D_T\nabla^2 T\label{eq4}
    \end{equation}
and $D_T$ is the thermal diffusion coefficient. The systems with
very short relaxation time have very short memory. On the other
hand for $\tau\to \infty$ Eq.~(\ref{eq3}) has the form of the
thermal wave (undamped) equation, or \textit{ballistic} thermal
equation. In the solid state physics the \textit{ballistic}
phonons or electrons are those for which $\tau\to \infty$. The
experiments with \textit{ballistic} phonons or electrons
demonstrate the existence of the \textit{wave motion} on the
lattice scale or on the electron gas scale.
    \begin{equation}
    \frac{\partial^2 T}{\partial t^2}=\frac{D_T}{\tau}\nabla^2T.\label{eq5}
    \end{equation}
For the systems with very long memory Eq.~(\ref{eq3}) is time symmetric equation with no arrow of time, for the Eq.~(\ref{eq5}) does not change the shape when $t\to -t$.

In Eq.~(\ref{eq3}) we define:
    \begin{equation}
    v=\left(\frac{D_T}{\tau}\right)^{\frac12},\label{eq6}
    \end{equation}
    velocity of thermal wave propagation and
    \begin{equation}
    \lambda=v\tau\label{eq7}
    \end{equation}
    where $\lambda$ is the mean free path of the heat carriers. With formula~(\ref{eq6}) equation (\ref{eq3}) can be written as
    \begin{equation}
    \frac{1}{v^2}\frac{\partial^2T}{\partial t^2}+\frac{1}{\tau v^2}\frac{\partial T}{\partial t}=\nabla^2 T.\label{eq8}
    \end{equation}
\subsection{Damped wave equation, thermal carriers in potential well, $V$}
From the mathematical point of view equation:
    $$
    \frac{1}{v^2}\frac{\partial^2 T}{\partial t^2}+\frac{1}{D}\frac{\partial T}{\partial t}=\nabla^2 T
$$
is the hyperbolic partial differential equation (PDE). On the other hand Fourier equation
    \begin{equation}
    \frac{1}{D}\frac{\partial T}{\partial t}=\nabla^2 T\label{eq9}
    \end{equation}
    and Schr\"{o}dinger equation
    \begin{equation}
    i\hbar\frac{\partial \Psi}{\partial t}=-\frac{\hbar^2}{2m}\nabla^2 \Psi\label{eq10}
    \end{equation}
    are the parabolic equations. Formally with substitutions
    \begin{equation}\label{eq11}
    T\leftrightarrow\Psi, \quad t\leftrightarrow it
    \end{equation}
Fourier equation~(\ref{eq9}) can be written as
    \begin{equation}\label{eq12}
  i\hbar\frac{\partial\Psi}{\partial t}=-D\hbar\nabla^2\Psi
    \end{equation}
and by comparison with Schr\"{o}dinger equation one obtains
\begin{equation}\label{eq13}
  D_T\hbar=\frac{\hbar^2}{2m}
\end{equation}
and

\begin{equation}\label{eq14}
  D_T=\frac{\hbar}{2m}.
    \end{equation}
Considering that $D_T=\tau v^2$(\ref{eq6}) we obtain from~(\ref{eq14})
    \begin{equation}\label{eq15}
  \tau=\frac{\hbar}{2mv^2}.
    \end{equation}
Formula~(\ref{eq15}) describes the relaxation time for quantum thermal precesses.

Starting with Schr\"{o}dinger equation for particle with mass~$m$ in potential~$V$:
    \begin{equation}\label{eq16}
  i\hbar\frac{\partial \Psi}{\partial t}=-\frac{\hbar^2}{2m}\nabla^2\Psi+V\Psi
    \end{equation}
and performing the substitution~(\ref{eq11}) one obtains
    \begin{equation}
    \hbar\frac{\partial T}{\partial t}=\frac{\hbar^2}{2m}\nabla^2T-VT\label{eq17}
    \end{equation}
and
    \begin{equation}\label{eq18}
    \frac{\partial T}{\partial t}=\frac{\hbar}{2m}\nabla^2T-\frac{V}{\hbar}T.
    \end{equation}
Equation~(\ref{eq18}) is Fourier equation (parabolic PDE) for $\tau=0$. For $\tau\neq0$ we obtain
    \begin{eqnarray}
    \tau\frac{\partial^2 T}{\partial t^2}+\frac{\partial T}{\partial t}+\frac{V}
    {\hbar}T&=&\frac{\hbar}{2m}\nabla^2 T,\label{eq19}\\
    \tau&=&\frac{\hbar}{2mv^2}\label{eq20}
    \end{eqnarray}
    or
$$
 \frac{1}{v^2}\frac{\partial^2 T}{\partial t^2}+\frac{2m}{\hbar}\frac{\partial T}{\partial t}+\frac{2mV}{\hbar^2}T=
 \nabla^2T.
$$
\section{Relaxation, Schr\"{o}dinger equation, strings}
\subsection{Model Schr\"{o}dinger equation}
With the substitution~(\ref{eq11}) equation~(\ref{eq19}) can be written as

\begin{equation}\label{eq21}
  i\hbar\frac{\partial\Psi}{\partial t}=V\Psi-\frac{\hbar^2}{2m}\nabla^2\Psi-\tau\hbar\frac{\partial^2\Psi}{\partial t^2}.
\end{equation}
  The new term, relaxation term
  \begin{equation}
  \tau\hbar\frac{\partial\Psi}{\partial t^2}\label{eq22}
  \end{equation}
  describes the interaction of the particle with mass $m$ with space-time. The relaxation time $\tau$ can be calculated as:
    \begin{equation}
    \frac{1}{\tau}=\frac{1}{\tau_{e-p}}+\cdots+\frac{1}{\tau_{\rm Planck}}\label{eq23}
    \end{equation}
where, for example $\tau_{\rm e-p}$ denotes the scattering of the
particle $m$ on the electron-positron pair $(\tau_{\rm e-p}\sim
10^{-17}~{\rm s})$  and the shortest relaxation time $\tau_{\rm
Planck}$ is the Planck time $(\tau_{\rm Planck}\sim10^{-43}~{\rm
s})$.

From equation~(\ref{eq23}) we conclude that $\tau\approx\tau_{\rm Planck}$ and equation~(\ref{eq21}) can be written as
    \begin{equation}
    i\hbar\frac{\partial \Psi}{\partial t}=
    V\Psi-\frac{\hbar^2}{2m}\nabla^2\Psi-\tau_{\rm Planck}\hbar\frac{\partial^2\Psi}{\partial t^2}\label{eq24}
\end{equation}
where
    \begin{equation}
    \tau_{\rm Planck}=\frac12\left(\frac{\hbar G }{c^5}\right)^{\frac12}=\frac{\hbar}{2M_Pc^2}.\label{eq25}
    \end{equation}
In formula~(\ref{eq25}) $M_P$ is the mass Planck. Considering Eq.~(\ref{eq25}), Eq.~(\ref{eq24}) can be written as
    \begin{equation}
    i\hbar\frac{\partial \Psi}{\partial t}=-\frac{\hbar^2}{2m}\nabla^2\Psi+V\Psi-\frac{\hbar^2}{2M_P}\nabla^2\Psi+
\frac{\hbar^2}{2M_P}\nabla^2\Psi-\frac{\hbar^2}{2M_Pc^2}\frac{\partial^2\Psi}{\partial t^2}.\label{eq26}
    \end{equation}
    The last two terms in Eq.~(\ref{eq6}) can be defined as the \textit{Bohmian} pilot wave
        \begin{equation}
        \frac{\hbar^2}{2M_P}\nabla^2\Psi-\frac{\hbar^2}{2M_Pc^2}\frac{\partial^2\Psi}{\partial t^2}=0\label{eq27}
    \end{equation}
i.e.
    \begin{equation}
    \nabla^2\Psi-\frac{1}{c^2}\frac{\partial^2\Psi}{\partial t^2}=0.\label{eq28}
    \end{equation}
It is interesting to observe that pilot wave $\Psi$ do not depends on the mass of the particle.
With postulate~(\ref{eq28}) we obtain from equation~(\ref{eq26})
    \begin{equation}
    i\hbar\frac{\partial \Psi}{\partial t}=-\frac{\hbar^2}{2m}\nabla^2\Psi+V\Psi-
    \frac{\hbar^2}{2M_P}\nabla^2\Psi\label{eq29}
    \end{equation}
    and simultaneously
    \begin{equation}
    \frac{\hbar^2}{2M_P}\nabla^2\Psi-\frac{\hbar^2}{2M_Pc^2}\frac{\partial^2\Psi}{\partial t^2}=0\label{eq30}
    \end{equation}
    In the operator form Eqs.~(\ref{eq9}) and (\ref{eq20}) can be written as
    \begin{equation}
    \hat{E}=\frac{\hat{p}^2}{2m}+\frac{1}{2M_Pc^2}\hat{E}^2,\label{eq31}
    \end{equation}
    where $\hat{E}$ and $\hat{p}$ denotes the operator for energy and
    momentum of the particle with mass $m$.
    Equation~(\ref{eq31}) is the new dispersion relation for quantum particle with mass $m$. From Eq.~(\ref{eq21}) one
    can concludes that Schr\"{o}dinger quantum mechanics is valid for particles with mass $m\ll M_P$.
    But pilot wave $\Psi$ exist independent of the mass of the particles.

For particles with mass $m\ll M_P$ Eq.~(\ref{eq9}) has the form
    \begin{equation}
    i\hbar\frac{\partial \Psi}{\partial t}=-\frac{\hbar^2}{2m}\nabla^2\Psi+V\Psi.\label{eq32}
    \end{equation}
    \subsection{Schr\"{o}dinger equation and the the strings}
In the case when $m\approx M_P$ Eq.~(\ref{eq29}) can be written as
    \begin{equation}
    i\hbar\frac{\partial \Psi}{\partial t}=-\frac{\hbar^2}{2M_P}\nabla^2\Psi+V\Psi\label{eq33}
    \end{equation}
    but considering Eq.~(\ref{eq30}) one obtains
    \begin{equation}
    i\hbar\frac{\partial \Psi}{\partial t}=-\frac{\hbar^2}{2M_Pc^2}
    \frac{\partial^2\Psi}{\partial t^2}+V\Psi\label{eq34}
\end{equation}
or
    \begin{equation}
    \frac{\hbar^2}{2M_Pc^2}\frac{\partial^2\Psi}
{\partial t^2}+i\hbar\frac{\partial \Psi}{\partial t}-V\Psi=0.\label{eq35}
    \end{equation}
We look for the solution of Eq.~(\ref{eq35}) in the form
    \begin{equation}
    \Psi(x,t)=e^{i\omega t}u(x).\label{eq36}
    \end{equation}
    After substitution formula~(\ref{eq16}) to Eq.~(\ref{eq35}) we obtain
    \begin{equation}
    \frac{\hbar^2}{2M_Pc^2}\omega^2+\omega\hbar+V(x)=0\label{eq37}
    \end{equation}
with the solution \jot=0.8cm
    \begin{eqnarray}
    \omega_1&=&\frac{-M_Pc^2+M_Pc^2\sqrt{1-\frac{2V}{M_Pc^2}}}{\hbar}\label{eq38}\\
    \omega_2&=&\frac{-M_Pc^2-M_Pc^2\sqrt{1-\frac{2V}{M_Pc^2}}}{\hbar}\nonumber
    \end{eqnarray}
for $\frac{M_Pc^2}{2}>V$ and \jot=0.8cm
\begin{eqnarray}
    \omega_1&=&\frac{-M_Pc^2+iM_Pc^2\sqrt{\frac{2V}{M_Pc^2}-1}}{\hbar}\label{eq39}\\
    \omega_2&=&\frac{-M_Pc^2-iM_Pc^2\sqrt{\frac{2V}{M_Pc^2}-1}}{\hbar}\nonumber
    \end{eqnarray}
for $\frac{M_Pc^2}{2}<V$.

Both formulae(\ref{eq38}) and (\ref{eq39}) describe the string
oscillation, formula~(\ref{eq27}) damped oscillation and formula
(\ref{eq28}) overdamped string oscillation.

\section{Conclusion}
D.~Bohm presented the pilot wave theory in 1952, and de Broglie
had presented a similar theory in the mid 1920s. It was rejected
in the 1950s and the initial rejection had nothing to do with
Bohm's later work.

There is always the possibility that the pilot wave has a
primitive, mind like property. That's how Bohm described it. We
can say that all the particles in Universe and even Universe have
their own \textit{pilot waves}, their own information. In this
paper we discuss in very simple nonrelativistic way possible
extention of the Schr\"{o}dinger equation with relaxation process
included. As the first approximation this lead us to the inclusion
of the Planck time, i.e. gravity to the quantum description of the
processes in the space time. The relaxation time $\tau_{\rm
Planck}$ is the decoherence time~\cite{6} or the Ehrenfest
time~\cite{7} and describes the collapse of the pilot wave after
the interaction with the apparatus.


\newpage
\begin{thebibliography}{99}
\bibitem{1} Julian Barbour, \textit{The End of Time}, Oxford University Press, 2000.
\bibitem{2} Jeffrey F. Addicot, \textit{Ohio State Law Journal}, 2003.
\bibitem{3} M. Kozlowski, J. Marciak-Kozlowska, The time arrow in a Planck gas,
\textit{Foundations of Physics Letters}, \textbf{10}, p. 295, (1997).
\bibitem{4} M. Kozlowski, J. Marciak-Kozlowska, The smearing out of the thermal initial
conditions created in a planck Era, \textit{Foundation of Physics Letters}, \textbf{10}, p. 599, (1997).
\bibitem{5} M. Kozlowski, J. Marciak-Kozlowska, Klein-Gordon thermal equation for Planck gas,
\textit{Foundations of Physics Letters}, \textbf{12}, p. 93, (1999).
\bibitem{6} Tamas Geszti, http://lanl.arxiv.org/quant-ph/0401086.
\bibitem{7} G. P. Berman at al., http://lanl.arxiv.org/quant-ph/0401038.
\end{thebibliography}
\end{document}